\documentclass[12pt]{iopart}
\begin{document}
\title{Halos of Unified Dark Matter Scalar Field}

\author{Daniele Bertacca}
\address{Dipartimento di Fisica "Galileo Galilei", Universit\'a di Padova, 
and INFN Sezione di Padova, via F. Marzolo, 8 I-35131 Padova, Italy}
\ead{daniele.bertacca@pd.infn.it}
\author{Nicola Bartolo}
\address{Dipartimento di Fisica "Galileo Galilei", Universit\'a di Padova, 
and INFN Sezione di Padova, via F. Marzolo, 8 I-35131 Padova, Italy}
\ead{nicola.bartolo@pd.infn.it}
\author{Sabino Matarrese}
\address{Dipartimento di Fisica "Galileo Galilei", Universit\'a di Padova, 
and INFN Sezione di Padova, via F. Marzolo, 8 I-35131 Padova, Italy}
\ead{sabino.matarrese@pd.infn.it}
\date{\today}

\begin{abstract}
We investigate the static and spherically symmetric solutions
of Einstein's equations for a scalar field with non-canonical kinetic term, 
assumed to provide both the dark matter and dark energy components 
of the Universe. 
In particular, we give a prescription to obtain 
solutions (dark halos) whose rotation curve $v_c(r)$ is 
in good agreement with observational data.     
We show that there exist suitable scalar field Lagrangians 
that allow to describe the cosmological background evolution and 
the static solutions with a single dark fluid. 

\end{abstract}

\maketitle

\section{Introduction}
The confidence region of cosmological parameters emerging from the 
analysis of data from type Ia Supernovae (SNIa),
Cosmic Microwave Background (CMB) anisotropies and 
the large scale structure of the Universe, suggests that 
two dark components govern the dynamics of the present universe.
These components are the Dark Matter (DM), 
responsible for structure formation, and an additional Dark Energy (DE)
component that drives the cosmic acceleration observed at present.
In this paper we focus on Unified models of Dark Matter and dark energy 
(UDM) that can provide an alternative to our interpretation of 
the nature of the dark components of our universe.
These models have the advantage over the DM + DE models (e.g. 
$\Lambda$CDM) that one can describe the dynamics of the universe
with a single dark fluid which triggers the accelerated expansion 
at late times and is also the one which has to cluster in order to 
produce the structures we see today. 
However, the viability of UDM models strongly depends on the 
value of the effective speed of sound $c_s$ 
\cite{Huss,Garriga:1999vw,Mukhanov:2005sc}, which has to be small enough 
to allow structure formation  
\cite{Sandvik:2002jz,Giannakis-Hu,Bertacca:2007cv} and   
to reproduce the observed pattern of 
CMB temperature anisotropies 
\cite{Huss,Bertacca:2007ux,Sandvik:2002jz,Finelli}.\\

Several adiabatic or, equivalently,
purely kinetic models have been investigated in the literature.   
For example, the generalized 
Chaplygin gas (\cite{Kamenshchik:2001cp,Bilic02,Bento:2002ps} (see also
\cite{GCG}), the Scherrer \cite{Scherrer:2004au} 
and generalized Scherrer \cite{Bertacca:2007ux} solutions, 
the single dark perfect fluid with a simple 2-parameter barotropic 
equation of state \cite{Bruni}, or the homogeneous scalar field deduced 
from the galactic halo space-times \cite{DiezTejedor:2006qh}.\\ 

Moreover, one can build up scalar field models for which the 
constraint that the Lagrangian is constant along the classical trajectories 
allows to describe a UDM fluid 
\cite{Bertacca:2007ux} (see also Ref.~\cite{BI}, for a different approach). 
Alternative approaches to the unification of DM 
and DE have been proposed in Ref.~\cite{takayana}, 
in the frame of supersymmetry, and in Ref.~\cite{bono}, 
in connection with the solution of the strong CP problem. \\

One could easily reinterpret UDM models based on a scalar field Lagrangian 
in terms of -- generally non-adiabatic -- fluids 
\cite{DiezTejedor:2005fz,Brown:1992kc}. 

A complete analysis of UDM models should necessarily include 
the study of static solutions of Einstein's field
equations. This is complementary to the study of cosmological background
solutions and would allow to impose further constraints to the Lagrangian 
of UDM models.
The authors of Refs.~\cite{ArmendarizPicon:2005nz} and 
\cite{DiezTejedor:2006qh} have studied spherically symmetric and 
static configuration for k-essence models. 
In particular, they studied models where the rotation velocity becomes flat 
(at least) at large radii of the halo.
In these models the scalar field pressure is not small 
compared to the mass-energy density, similarly to what found in the study 
of general fluids in
Refs.~\cite{Bharadwaj:2003iw,Lake:2003tr,Faber:2005xc,Faber:2006sb}, 
and the Einstein's equations of motion do not reduce to the equations 
of Newtonian gravity. 
Further alternative models have been considered, even 
with a canonical kinetic term in the Lagrangian, that
describe dark matter halos in terms of bosonic scalar fields,
see e.g. Refs. 
\cite{Lee:1995af,Wetterich,Arbey:2001jj,Arbey:2003sj,Bilic:2005sn}.\\
In this paper we assume that our scalar field 
configurations only depend on the radial direction.  
Three main results are achieved. First, we are able to find a purely kinetic
Lagrangian which allows simultaneously to provide a flat rotation curve and
to realize a unified model of dark matter and dark energy 
on cosmological scales.
Second, we have found an invariance property of the expression for the halo 
rotation curve. This allows to find purely kinetic Lagrangians that reproduce
the same rotation curves that are obtained starting from a given 
density profile 
within the standard Cold Dark Matter (CDM) paradigm. 
Finally, we consider a more general class 
of models with non-purely kinetic Lagrangians. In this case we extend to 
the static and spherically symmetric space-time metric the procedure used 
in Ref.~\cite{Bertacca:2007ux} to find UDM solutions in a cosmological setting.
Such a procedure requires that the Lagrangain is constant 
along the classical trajectories; we are thus able 
to provide the conditions to obtain reasonable rotation curves 
within a UDM model
of the type discussed in Ref.~\cite{Bertacca:2007ux}. 

The plan of the paper is as follows. In Section 2 we provide the general 
framework for the study of static spherically symmetric solutions in 
UDM models. Section 3 is devoted to the general 
analysis of purely kinetic Lagrangians, while in Section 4 we analyze 
more general models with non-canonical kinetic term. Our main conclusions 
are drawn in Section 5. In the Appendix, for completeness we provide the   
spherical collapse top-hat solution for UDM models based on 
purely kinetic scalar-field Lagrangians, which allow to connect the 
cosmological solutions to the static configurations studied here.

\section{Static solutions in Unified Dark Matter scalar field models }

We consider the following action 
\begin{equation}\label{eq:action}
S = S_{G} + S_{\varphi}+S_b = 
\int d^4 x \sqrt{-g} \left[\frac{R}{2}+\mathcal{L}(\varphi,X)\right]+S_b
\end{equation}
where $S_b$ describes the baryonic matter and 
\begin{equation}\label{x}
X=-\frac{1}{2}\nabla_\mu \varphi \nabla^\mu \varphi \;.
\end{equation}
We use units such that $8\pi G = c= 1$ and signature $(-,+,+,+)$; greek 
indices run over space-time dimensions, while latin indices label spatial 
coordinates.\\
The energy-momentum tensor of the scalar field $\varphi$ is
\begin{equation}
  \label{energy-momentum-tensor}
  T^{\varphi}_{\mu \nu } = 
- \frac{2}{\sqrt{-g}}\frac{\delta S_{\varphi }}{\delta
    g^{\mu \nu }}=\frac{\partial \mathcal{L}
(\varphi ,X)}{\partial X}\nabla _{\mu }\varphi
  \nabla _{\nu }\varphi +\mathcal{L}(\varphi ,X)g_{\mu \nu }\, ,
\end{equation}
and its equation of motion reads 
\begin{equation}
\label{eq-motion}
\nabla^\mu \left[\frac{\partial\mathcal{L}}{\partial(\partial_\mu \varphi)}
\right]=\frac{\partial\mathcal{L}}{\partial \varphi}\;. 
\end{equation}
We consider a scalar field which is static and spatially inhomogeneous, i.e.
such that $X<0$. 
In this situation the energy-momentum tensor is not described
by a perfect fluid and its stress energy-momentum tensor reads 
\begin{equation}
T^{\varphi}_{\mu \nu } = (p_\parallel+\rho) n_\mu n_\nu - \rho g_{\mu \nu }
\end{equation}
where 
\begin{equation}
\label{energy-density}
\rho=-p_{\perp}=-\mathcal{L}\;,
\end{equation}
$n_\mu=\nabla _{\mu }\varphi/\sqrt{-2X}$
and $p_{\parallel}=\mathcal{L} - 2X\partial \mathcal{L}/\partial X $.
In particular, $p_{\parallel}$ is the pressure in the direction parallel to 
$n_\mu$ whereas $p_{\perp}$ is the pressure in the direction orthogonal to 
$n_\mu$. It is simpler to work with a new definition of $X$. 
Indeed, defining $X=-\chi$ we have
\begin{eqnarray}
\label{def_n_with_chi}
n_\mu=\nabla _{\mu }\varphi/(2\chi)^{1/2}\\ 
\label{def-p_with_chi}
p_{\parallel}=2\chi\frac{\partial\rho}{\partial\chi} - \rho \;. 
\end{eqnarray}
Let us consider for simplicity the general static spherically 
symmetric space-time metric i.e.
\begin{equation}
ds^2=-\exp{(2\alpha(r))}\;dt^2+\exp{(2\beta(r))}\;dr^2+r^2 d\Omega^2\, ,
\end{equation}
where $d\Omega^2=d\theta^2+\sin^2 \theta d\phi^2$ and $\alpha$
and $\beta$ are two functions that only depend upon $r$.\\
As the authors of Refs.~\cite{ArmendarizPicon:2005nz,DiezTejedor:2006qh} have 
shown, it is easy to see that the non-diagonal term $T^{rt}$ vanishes. 
Therefore $\varphi$ could be either 
strictly static or depend only on time. In this paper we study the
solutions where $\varphi$ depends on the radius only.\\

In the following we will consider some cases where the 
baryonic content is not negligible in the halo. In this case  
we will assume that most of the baryons are concentrated 
within a radius $r_b$. If we define $M_*$ as the entire mass 
of the baryonic component then for $r>r_b$ we can simply assume that $M_*$ 
is concentrated in the center of the halo. 

Considering, therefore, the halo for $r>r_b$,
starting from the Einstein's equations and the covariant conservation of
the stress-energy (or from the equation of
motion of the scalar field, Eq.~(\ref{eq-motion})), 
we obtain 
\begin{eqnarray}
\label{eqE1}
\fl
\frac{1}{r^2}\left\{1-\left[r\exp{\left(-2\beta\right)}\right]'\right\}=\rho
&\Longleftrightarrow& \frac{dM}{dr}=4\pi\rho r^2 \;,\\
\label{eqE2}
\fl
\frac{1}{r^2}\left\{\exp{\left[-2(\alpha+\beta)\right]}
\left[r\exp{\left(2\alpha\right)}\right]'-1\right\}=p_{\parallel}
&\Longleftrightarrow& \alpha'=
\frac{\frac{M+M_*}{8\pi}+\frac{p_{\parallel}r^3}{2}}
{r^2\left[1-\frac{M+M_*}{4\pi r}\right]} \;, \\
\label{eqE3}
\fl
\frac{\exp{\left[-(\alpha+2\beta)\right]}}{r}
\left\{\left[r\exp{\alpha}\right]'\beta'-
\left[r\left(\exp{\alpha}\right)'\right]'\right\}=\rho \;,
\end{eqnarray}
\begin{equation}
\label{eq-conserv-st}
\fl
\frac{dp_\parallel}{dR}=-(p_\parallel+\rho)
\end{equation}
\noindent
(which are the $00$, $rr$ and $\theta\theta$ components of Einstein's 
equations and the $r$ component of the continuity equation respectively)
where $\exp{(-2\beta(r))}=1-(M+M_*)/(4\pi r)$ and
$R=\ln[r^2\exp(\alpha(r))]$. Here a prime indicates differentiation with 
respect to the radius $r$.\\ 
A first comment is in order here. If {\it i)} 
$\beta'=0$ and {\it ii)} $\left[r\left(\exp{\alpha}\right)'\right]'>0$,  
then we can immediately see that $\rho<0$. These conditions must therefore 
be avoided when trying to find a reasonable rotation curve. For example,
neglecting the baryonic mass, the special case of $\rho=A/r^2$ and 
$\exp(\alpha)\sim r^m$, where $A$ and $m$ are constants, fall into this case.
We thus recover the \emph{no-go} theorem derived in 
Ref.~\cite{DiezTejedor:2006qh} under the assumption  
that the rotation curve $v_c\ll 1$ is constant for all $r$.\\ 

The value of the circular velocity $v_c$ is determined by the assumption 
that a massive test particle is located at $\theta=\pi/2$. 
We define as massive test particle the object that
sends out a luminous signal to the observer who is considered to be 
stationary and far away from the halo. In this case the value of $v_c=r\phi'$ 
can be rewritten as 
\begin{equation}
\label{v_c}
v_c^2=\frac{p_\parallel r^2/2+(M+M_*)/(8\pi r)}
{1-\left[p_\parallel r^2/2+3(M+M_*)/(8\pi r)\right]} \;, 
\end{equation}
but when we consider the weak-field limit condition 
$(M+M_*)/(8\pi r)\ll 1$ and since the rotation velocities of the halo of a 
spiral galaxy are typically non-relativistic, $v_c \ll 1$, 
Eq.~(\ref{v_c}) simplifies to \cite{ArmendarizPicon:2005nz}
\begin{equation}
\label{v_c-2}
v_c^2 \approx \frac{M+M_*}{8\pi r}+\frac{p_\parallel r^2}{2}\;.
\end{equation}

A second comment follows from the fact that the pressure is not small
compared to the mass-energy density. In other words we do not require
that general relativity reduces to Newtonian gravity (see also 
Refs.~\cite{Bharadwaj:2003iw,Lake:2003tr,Faber:2005xc,Faber:2006sb}). 
Notice also that in the region where 
$v_c \approx {\rm const.} \ll 1$ it is easy to see that in general 
$\exp(\alpha) \approx {\rm const.}$ since from Eqs.~(\ref{eqE2}) 
and~(\ref{v_c-2}) one obtains $r \alpha' \approx v_c^2$.
 
Finally, let us point out one of our main results. 
We can see that the relation (\ref{v_c-2}) 
is invariant under the following transformation
\begin{equation}
\label{transformation}
\rho\longrightarrow\widetilde{\rho}=\rho+\sigma(r) \quad\quad\quad
p_\parallel\longrightarrow\widetilde{p_\parallel}=p_\parallel+q(r)
\end{equation}
if  
\begin{equation}
\label{cond-q-sigma}
3q(r)+rq(r)'=-\sigma(r)\;,
\end{equation}
up to a proper choice of some integration constants.
Thanks to this transformation we can consider an ensemble of solutions that 
have the same rotation curve. We will come back to this point in more 
detail in 
the next section. Obviously, these solutions have to
satisfy the Einstein's equations (\ref{eqE1}), (\ref{eqE2}) and (\ref{eqE3}),
and the covariant conservation of the stress-energy 
(\ref{eq-conserv-st}). 
Moreover, we will require the validity of the weak energy conditions, 
$\rho \geq 0$ and $p_\parallel+\rho \ge 0$, i.e. 
\begin{equation}
2\frac{\exp{\left(-2\beta\right)}}{r}(\alpha'+\beta')=
 2\chi\frac{\partial\rho}{\partial\chi}\ge 0\;.
\end{equation}

In the following  sections we will consider first a purely kinetic 
Lagrangian $\mathcal{L}(X)$ and then two Lagrangians 
$\mathcal{L}=f(\varphi)g(X)$ and $\mathcal{L}=g(X)-V(\varphi)$.  

\section{Unified Dark Matter models with purely kinetic Lagrangians}

Let us consider a scalar field Lagrangian $\mathcal{L}$ 
with a non-canonical kinetic term that depends only on $X$ or $\chi$.   
Moreover, in this section we assume that 
$M_*=0$ (or $M\gg M_*$).

First of all we must impose that $\mathcal{L}$ is negative when $X<0$, so that
the energy density is positive. Therefore, we define a new positive function 
\begin{equation}
g_s(\chi)\equiv - \mathcal{L}(X)\;.
\end{equation}
As already explained in Ref.~\cite{ArmendarizPicon:2005nz}, when
the equation of state $p_\parallel=p_\parallel(\rho)$ is known, one 
can write the purely kinetic Lagrangian that describes this dark fluid 
with the help of Eqs.~(\ref{energy-density}) and 
(\ref{def-p_with_chi}). 
Alternatively, using (\ref{eq-conserv-st}), one can connect 
$p_\parallel$ and $\rho$ in terms of $r$ through the variable $R$. 
Moreover, it is easy to see that 
starting from the field equation of motion (\ref{eq-motion}), 
there exists another relation that connects $\chi$ (i.e. $X$) with $r$.
This relation is  
\begin{equation}
\label{gs-chi-r}
\chi \left[\frac{dg_s(\chi)}{d\chi}\right]^2=
\frac{k}{\left[r^2\exp{\alpha(r)}\right]^2}
\end{equation}
with $k$ a positive constant. 
If we add an additive constant to $g_s(\chi)$, 
the solution (\ref{gs-chi-r}) remains unchanged.
One can see this also through the Eq.~(\ref{eq-conserv-st}).
Indeed, using Eqs.~(\ref{energy-density}) and ~(\ref{def-p_with_chi})
one immediately finds that Eq.~(\ref{eq-conserv-st}) is invariant under
the transformation $\rho \rightarrow \rho + K$ \;
$p_\parallel \rightarrow p_\parallel - K$.
In this way we can add the cosmological constant
$K=\Lambda$ to the Lagrangian and we can describe the dark matter and the
cosmological constant-like dark energy as a single dark fluid i.e. 
as Unified Dark Matter (UDM). 

Let us notice that one can adopt two approaches to find 
reasonable rotation curves $v_c(r)$. A static solution  can be studied
in two possible ways: 
\begin{itemize} 
\item[i)]
The first approach consists simply in adopting directly
a Langrangian that provides a viable cosmological  
UDM model and exploring what 
are the conditions under which it can give a static solution 
with a rotation curve that is flat at large radii. This prescription has been 
already applied, for example, in Ref.~\cite{ArmendarizPicon:2005nz}.
\item[ii)]
A second approach consists in exploiting the invariance 
property of Eq.~(\ref{v_c-2}), 
with respect to the transformation (\ref{transformation}) (when the condition
(\ref{cond-q-sigma}) is satisfied). 
Usually in the literature one reduces the problem to the 
Newtonian gravity limit, because one makes use of a CDM density profile, 
i.e. one assumes that in 
Eq.~(\ref{v_c-2}), $p_\parallel \ll M/(4\pi r^3)$. We can therefore use 
Eqs.~(\ref{transformation}) and (\ref{cond-q-sigma})
to obtain energy density and pressure profiles $\rho(r)$ and 
$p_\parallel(r)$ that reproduce the same rotation curve in a model with 
non-negligible pressure. 
Next, we find an acceptable equation of state $p_\parallel=p_\parallel(\rho)$
such that we can reconstruct, through Eqs.~(\ref{energy-density}) and 
(\ref{def-p_with_chi}), the expression for the Lagrangian $\mathcal{L}$. 
Such a procedure establishes a mapping between UDM and CDM solutions 
that predict the same halo rotation curve $v_c(r)$.
As a starting point we could, of course, use very different CDM density 
profiles to this aim, such as the modified isothermal-law profile
\cite{King:1962wi}, the Burkert profile \cite{Burkert:1995yz}, the Moore
profile \cite{Moore:1999gc}, the Navarro-Frenk-White profile 
\cite{Navarro:1996gj,Navarro:2003ew} or the profile proposed 
by Salucci et al. (see for example \cite{Salucci}). \\
As we have already mentioned, the possible solutions one finds in 
this way have to
satisfy the Einstein equations (\ref{eqE1}), (\ref{eqE2}) and (\ref{eqE3}),
the conservation of stress-energy (\ref{eq-conserv-st}) and the weak energy 
conditions. Moreover, the resulting UDM scalar field Lagrangian must 
be able to provide cosmological solutions that yield an acceptable 
description of the cosmological background 
(see, e.g., Ref.~\cite{Bertacca:2007ux}) and low effective speed of sound 
(see for example Refs.~\cite{Garriga:1999vw,Mukhanov:2005sc,Bertacca:2007cv}) 
so that cosmic structure formation successfully takes place and 
CMB anisotropies fit the observed pattern 
\cite{Sandvik:2002jz,Giannakis-Hu,Finelli}. 
\end{itemize}

Below, using approach i), we provide a worked example of a UDM model
with purely kinetic Lagrangian which is able to describe a flat halo 
rotation curve and then, using approach ii), we give a general 
systematic procedure to obtain a possible Lagrangian of UDM model 
starting from a given CDM density profile.

\subsection{Approach i): The generalized Scherrer solution}

Let us consider the generalized Scherrer solution models obtained in 
Ref.~\cite{Bertacca:2007ux}. These models are described by
the following Lagrangian
\begin{equation}
\mathcal{L}= -\Lambda + g_n \left(X-\hat{X}\right)^n 
\end{equation}
where $g_n>0$ is a suitable constant and $n>1$. 
The case $n=2$ corresponds to the unified model proposed by Scherrer
~\cite{Scherrer:2004au}.  
If we impose that today $[(X-\hat{X})/\hat{X}]^n \ll 1$, 
the background energy density can be written as 
\begin{equation}
\label{rho_lambda_dm}
\rho(a(t)) = \rho_\Lambda + \rho_{\rm DM} \;, 
\end{equation}
where $\rho_\Lambda$ behaves like a ``dark energy'' component 
($\rho_\Lambda = {\rm const.}$) and 
$\rho_{\rm DM}$ behaves like a ``dark matter'' component 
i.e. $\rho_{\rm DM}\propto a^{-3}$, with $a(t)$ the scale factor.\\
A static solution for the generalized Scherrer model can be obtained 
in two possible ways:
\begin{itemize} 
\item[1)] Starting from the analysis of Ref.~\cite{DiezTejedor:2006qh}, 
in the case of a barotropic Lagrangian for the homogeneous field.
The authors of Ref.~\cite{DiezTejedor:2006qh} indeed concluded that 
for $n \gg 1$ flat halo rotation curves can be obtained. 
In particular they studied spherically symmetric 
solutions with the following metric, 
\begin{equation}
ds^{2}=-\left(\frac{r}{r_{\star}}\right)^{l}dt^{2}+N(r)dr^{2}+
r^{2}d\Omega^{2}.\label{eq:flat}
\end{equation}
where $r_{\star}$ is a suitable length-scale and $l=2v_{c}^2$.
In the trivial case where $N(r)$ is constant they find $\mathcal{L}(X)\propto 
X^{2/l}$ with $l \ll 1$.
For $X \gg \hat{X} $ the Lagrangian 
$\mathcal{L}= -\Lambda + g_n (X-\hat{X})^n $ takes precisely this form.
\item[2)] In the analysis of Ref.~\cite{ArmendarizPicon:2005nz},  
solutions where $\varphi$ is only a function of the radius are considered. 
When the Lagrangian has the form $\mathcal{L}\propto X^n $, with $n \sim 10^6$ 
the halo rotation curve becomes flat at large radii. 
In this case $n$ must be an odd natural number, such that the energy density 
is positive. Our model is able to reproduce this situation when the matter 
density is large, i.e. when $|X| \gg \hat{X}$. 
\end{itemize}
Alternatively, if we wish to avoid large $n$ (c.f. case 2) above)
we can start from the following Lagrangian
\begin{equation}
\label{L-purelykessence}
\mathcal{L}= -\Lambda + \epsilon_X g_n  \left(|X|-\hat{X}\right)^n 
\end{equation}
where $\epsilon_X$ is some differentiable function of $X$ that is  
$1$ when $X\ge\hat{X}$ and $-1$ when $X\le-\hat{X}<0$. 
In this way when $X>\hat{X}>0$ we recover the Lagrangian of the generalized 
Scherrer solutions. When $X<0$ and $\chi=-X>\hat{X}$ we get
\begin{equation}
\mathcal{L}= -\Lambda - g_n \left(\chi-\hat{X}\right)^n 
\end{equation}
and, with the help of Eqs.~(\ref{energy-density}) and 
(\ref{def-p_with_chi}), we obtain 
\begin{equation}
\fl
\rho=-p_{\perp}=-\mathcal{L}\;, \quad \quad \quad
p_{\parallel}=(2n-1)g_n \left(\chi-\hat{X}\right)^n+2ng_n\hat{X} 
\left(\chi-\hat{X}\right)^{n-1}-
\Lambda\;.
\end{equation}
Now, requiring that $\chi$ be close to $\hat{X}$ (i.e.
$\left(\chi-\hat{X}\right)\ll \hat{X}$) and
$2ng_n\hat{X}\left(\chi-\hat{X}\right)^{n-1} \gg O(\Lambda)$,
and starting from the relation (\ref{gs-chi-r}) 
that connects $\chi$ with $r$, we get
\begin{equation}
\label{gs-chi-r-scherrer}
\left(\chi-\hat{X}\right)^{n-1}=\frac{k^{1/2}}{n g_n \hat{X}^{1/2}}\;
\frac{1}{r^2\exp{(\alpha(r))}}\;.
\end{equation}
Consistency with our approximations implies that we have to consider 
the following expressions for radial configurations with 
$r$ bigger than a minimum radius $r_{min}$. 
In this case  $p_{\parallel}$ and $\rho$ become
\begin{equation}
\fl
p_{\parallel}=\frac{A}{r^2\exp{(\alpha(r))}} \;, \quad \quad \quad
\rho=\frac{B}{[r^2\exp{(\alpha(r))}]^{n/(n-1)}}
\end{equation}
where $A=2(k\hat{X})^{1/2}$ and 
$B=g_n \left[k^{1/2}/(n g_n \hat{X}^{1/2})\right]^{n/(n-1)}$.\\
Using the Eqs.~(\ref{eqE1})~and~(\ref{eqE2}), we be able
to calculate the values of the metric terms $\exp{(\alpha)}$ and $\exp{(\beta)}$
 and, thus the value of $\rho$ and $p_{\parallel}$. Alternatively we know that 
when $v_c\approx {\rm const. \ll 1}$ at large radii, in a first approximation,
we can set $\exp(\alpha(r))\approx 
C = $ const.  Therefore for $n \neq 3$, we can write 
the function $M$ as 
\begin{equation}
M(r)\approx \frac{4\pi B}{C^{n/(n-1)}}
\left(\frac{n-1}{n-3}r^{\frac{n-3}{n-1}}+D\right)
\end{equation}
where we could also set $D=0$ for $n>3$. 
Instead, when $1<n<3$, the second term has to be larger than the first one.

In these cases $v_{c}^2$ becomes 
\begin{equation}
v_{c}^2(r) \approx \frac{A}{2 C}+
\frac{B}{2\;C^{n/(n-1)}}\left(\frac{n-1}{n-3}
\frac{1}{r^{2/(n-1)}}+\frac{D}{r}\right)\;.
\end{equation}
For $n=3$ we have 
\begin{equation}
M(r)\approx \frac{4\pi B}{C^{3/2}}
\ln\left(\frac{r}{\bar{r}}\right)+M(\bar{r})
\end{equation}
where $r > \bar{r}$ and
\begin{equation}
v_{c}^2\approx \frac{A}{2 C}+
\frac{B}{2\;C^{3/2}}\frac{1}{r}
\ln\left(\frac{r}{\bar{r}}\right)+\frac{M(\bar{r})}{8\pi r}\;.
\end{equation}
In other words we see that the circular velocity becomes  
approximately constant for sufficiently large $r$. 

However, let us stress that $\exp(\alpha(r))$ cannot be strictly  
constant, and that it should be chosen in such a way that  
the positivity of Eq.~(\ref{eqE3}) is ensured. 

This example can be generalized also to $M_*\neq 0$. Obviously,  
in such a case we have to assume that $r>r_b\ge r_{min}$. 
In this case $k$, $r_{min}$, $A$, $B$ (through 
$\exp (\beta(r))$) and $C$ depend on $M_*$.

The spherical top-hat solution for this model, which provides the link with 
the cosmological initial conditions, is described in the Appendix.  

\subsection{Approach ii): 
A general prescription to obtain UDM Lagrangians starting 
from a profile of an energy density distribution of CDM}

Defining the energy density distribution of CDM as
$\rho_{\rm CDM}(r)$ (with $p_{\rm CDM}=0$), the transformation 
(\ref{transformation})
 becomes 
\begin{equation}
\label{transformation-CDM}
\rho(r)=\rho_{\rm CDM}(r)+\sigma(r)\, , \quad\quad\quad
p_\parallel(r)=q(r) \;.
\end{equation}
Now, starting from a given CDM density profile, through 
Eqs.~(\ref{eqE1}), (\ref{eqE2}), 
(\ref{eq-conserv-st}) and   (\ref{cond-q-sigma})
we can determine~$\exp{(\alpha)}$~, ~$\exp{(\beta)}$, 
~$\rho$ ~and ~$p_\parallel$~. In a second step we provide the 
conditions to ensure that the energy density is positive 
\footnote{Thanks to this condition, through Einstein's Eq.~(\ref{eqE3}), 
we can evade the \emph{no-go} theorem derived in 
Ref.~\cite{DiezTejedor:2006qh}.}. 
In this case, after some simple but lenghty calculations, we find
\begin{eqnarray}
\label{Q}
\mathcal{Q}'(r)\left(r\frac{M_{\rm CDM}(r)}
{4\pi}-2r\mathcal{Q}(r)\right)
-2\mathcal{Q}^2(r)\nonumber \\ 
+\mathcal{Q}(r)\left(4r+3\frac{M_{\rm CDM}(r)}{4\pi}+
4r^3\rho_{\rm CDM}\right)=
\frac{r M_{\rm CDM}(r)}{4\pi}\left(4
+3r^2\rho_{\rm CDM}\right)\;,\\
\label{B}
\mathcal{B}(r)=\mathcal{Q}(r)-\frac{M_{\rm CDM}(r)}{4\pi}\;,\\
\label{A}
\mathcal{A}(r)=\frac{\mathcal{Q}(r)+\mathcal{B}(r)}{2\mathcal{B}(r)}\;,\\
\sigma(r)=\frac{1-\mathcal{Q}'(r)}{r^2}
\end{eqnarray}
\noindent
where $\mathcal{Q}(r) = r(r^2q+1)$~, ~$\mathcal{B}(r)=r\exp{(-2\beta)}$
and $\mathcal{A}(r)=(r\alpha'+1)$. Here we define
$M_{\rm CDM}(r)=4\pi \int_{0}^r \tilde{r}^2
\rho_{\rm CDM}(\tilde{r})~d\tilde{r}$.
At this point it is easy to see that Eq.~(\ref{Q}) does not admit a simple
analytical solution for a generic $\rho_{\rm CDM}$.
On the other hand we know that, through $\rho_{\rm CDM}$, all these functions
depend on the velocity rotation curve $v_c(r)$. Moreover $v_c^2(r)\ll 1$. 
Therefore, defining $\bar{v}_c$ as the value that $v_c$ assumes when the 
rotation curve is flat at large radii or the maximum value of $v_c$
with a particular profile of $\rho_{\rm CDM}$,
we can expand $\mathcal{Q}$, $\mathcal{A}$ and $\mathcal{B}$ 
as 
\begin{eqnarray}
\label{Q-pet}
\mathcal{Q}(r)=\mathcal{Q}_{(0)}(r)+\bar{v}_c^2 \mathcal{Q}_{(1)}(r)
+\frac{\left(\bar{v}_c^2\right)^2}{2!} \mathcal{Q}_{(2)}(r)+ \dots \;,
\nonumber\\
\label{A-pet}
\mathcal{A}(r)=\mathcal{A}_{(0)}(r)+\bar{v}_c^2 \mathcal{A}_{(1)}(r)
+\frac{\left(\bar{v}_c^2\right)^2}{2!} \mathcal{A}_{(2)}(r)+ \dots\;,
\nonumber\\
\label{B-pet}
\mathcal{B}(r)=\mathcal{B}_{(0)}(r)+\bar{v}_c^2 \mathcal{B}_{(1)}(r)
+\frac{\left(\bar{v}_c^2\right)^2}{2!} \mathcal{B}_{(2)}(r)+ \dots\;. 
\end{eqnarray}
\noindent

Following this procedure we can determine $\rho$ and $p_\parallel$ 
in a perturbative way, i.e.
\begin{equation}
\rho(r)=\rho_{(0)}(r)+\bar{v}_c^2 \rho_{(1)}(r) + 
\frac{\left(\bar{v}_c^2\right)^2}{2!} \rho_{(2)}(r)+ \dots\;,
\end{equation}
\begin{equation}
p_\parallel(r)=p_{\parallel \;(0)}(r)+\bar{v}_c^2 p_{\parallel \;(1)}(r)
+\frac{\left(\bar{v}_c^2\right)^2}{2!} p_{\parallel \;(2)}(r)+ \dots\;.
\end{equation}
\noindent
Now, looking at the various CDM density profiles
which have been proposed in the literature 
\cite{King:1962wi,Burkert:1995yz,Moore:1999gc,Navarro:1996gj,Navarro:2003ew,Salucci}, we see that we can always take $\rho_{\rm CDM}$
as
\begin{equation}
\rho_{\rm CDM}(r)=\bar{v}_c^2 \rho_{{\rm CDM}\;(1)}(r)\;, 
\end{equation}
then
\begin{equation}
 M_{\rm CDM}(r)=\bar{v}_c^2 M_{{\rm CDM}\;(1)}(r)=
 4\pi ~\bar{v}_c^2 \int_{0}^r \tilde{r}^2\rho_{{\rm CDM}\;(1)}
(\tilde{r})~d\tilde{r}\;.
\end{equation}
\noindent
For the zeroth-order terms we immediately obtain 
\begin{eqnarray}
\label{Q-0pet}
\mathcal{Q}_{(0)}=r\;,\nonumber\\
\label{A-0pet}
\mathcal{A}_{(0)}=1\;,\nonumber\\
\label{B-0pet}
\mathcal{B}_{(0)}=r\;.
\end{eqnarray}
\noindent
At the first order we get
\begin{eqnarray}
\label{Q-1pet}
\mathcal{Q}_{(1)}=\frac{2}{r}\int_{0}^r 
\tilde{r}^3\rho_{{\rm CDM}\;(1)}
(\tilde{r})~d\tilde{r}\;,\nonumber\\
\label{A-1pet}
\mathcal{A}_{(1)}=\frac{1}{2r} 
\frac{M_{{\rm CDM}\;(1)}(r)}{4\pi}\;,\nonumber\\
\label{B-1pet}
\mathcal{B}_{(1)}=\frac{2}{r}\int_{0}^r 
\tilde{r}^3\rho_{{\rm CDM}\;(1)}
(\tilde{r})~d\tilde{r}-\frac{M_{{\rm CDM}\;(1)}(r)}{4\pi} \;.
\end{eqnarray}
\noindent
For completeness we write also the second order for $\mathcal{Q}$
\begin{eqnarray}
\label{Q-2pet}
\mathcal{Q}_{(2)}=\frac{1}{r} \int_{0}^r d\check{r}
\frac{M_{{\rm CDM}\;(1)}(\check{r})}{4\pi}
\left[\frac{2}{\check{r}} \mathcal{Q}_{(1)}({\check{r}}) 
- {\check{r}}^2 \rho_{{\rm CDM}\;(1)}
({\check{r}}) \right] \,.
\end{eqnarray}
\noindent
Let us stress that if one considers also terms 
$O\left(\bar{v}_c^4\right)$, Eq.~(\ref{v_c})
instead of Eq.~(\ref{v_c-2}) should be used. 
In such a case, $v_c$ slightly changes
with respect to the velocity rotation curve that one obtains using a 
CDM density profile.

For our purposes we can consider only the zeroth and the first-order 
terms.
At this point, we can finally calculate the value of ~$\rho$ ~and
~$p_\parallel$~. We get
\begin{equation}
\fl
\rho(r)=\rho_{\rm CDM}(r)+\frac{1-\mathcal{Q}'(r)}{r^2}=
\bar{v}_c^2 \left(\frac{2}{r^4}\int_{0}^r 
\tilde{r}^3\rho_{{\rm CDM}\;(1)}
(\tilde{r})~d\tilde{r}-\rho_{\rm CDM\;(1)}(r)\right)\;,
\end{equation}
\begin{equation}
\fl
p_\parallel(r)=\frac{\mathcal{Q}(r)-r}{r^3}=\bar{v}_c^2
\frac{2}{r^4}\int_{0}^r \tilde{r}^3\rho_{{\rm CDM}\;(1)}
(\tilde{r})~d\tilde{r}\;.
\end{equation} 
\noindent
As far as the values of the metric terms $\exp{(\alpha)}$ and $\exp{(\beta)}$
are concerned, we obtain the following expressions
\begin{equation}
\fl
\exp{(2\alpha)}=\exp{(2\alpha(\hat{r}))}\exp{\left[\bar{v}_c^2\int_{\hat{r}}^r
\frac{1}{\tilde{r}^2}\frac{M_{{\rm CDM}\;(1)}
(\tilde{r})}{4\pi}
~d\tilde{r} \right]}
\end{equation} 
\begin{equation}
\fl
\exp{(-2\beta)}=1+\frac{\bar{v}_c^2}{r^2}\left(2\int_{0}^r 
\tilde{r}^3\rho_{{\rm CDM}\;(1)}
(\tilde{r})~d\tilde{r}-r \frac{M_{{\rm CDM}\;(1)}(r)}
{4\pi}\right)\;.
\end{equation}
\noindent
Now, it is immediate to see that if we want a positive energy density 
we have to impose $2~\int_{0}^r \tilde{r}^3\rho_{{\rm CDM}\;(1)}
(\tilde{r})~d\tilde{r} \ge r^4~\rho_{{\rm CDM}\;(1)}(r)$. From 
Eq.~(\ref{eqE1}) we know that $M(r)=4\pi \int_{\hat{r}_{(0)}}^r 
\tilde{r}^2\rho(\tilde{r})+ M(\hat{r}_{(0)})$ and
$M_{\rm CDM}(r)=4\pi \int_{\bar{r}}^r 
\tilde{r}^2\rho_{{\rm CDM}}(\tilde{r}) ~d\tilde{r}+M_{{\rm CDM}}(\bar{r})$. 
Therefore we need to know what is the relation 
between $\bar{r}$ and $\hat{r}_{(0)}$. 
This condition is easily obtained if we make use of 
Eq.~(\ref{v_c-2}). Indeed, we get
\begin{equation}\label{initial-cond}
\fl
\frac{M_{(1)}(\hat{r}_{(0)})-M_{{\rm CDM}\;(1)}(\bar{r})}{4 \pi}+
\frac{2}{\hat{r}_{(0)}}\int_{0}^{\hat{r}_{(0)}} 
\tilde{r}^3\rho_{{\rm CDM}\;(1)}(\tilde{r})~d\tilde{r}=
\int_{\bar{r}}^{\hat{r}_{(0)}}\tilde{r}^2
\rho_{{\rm CDM}\;(1)}(\tilde{r})~d\tilde{r}\; , 
\end{equation}
which finally guarantees the invariance of the rotation velocity with respect
to the transformation in Eqs.~(\ref{transformation}) and (\ref{cond-q-sigma}). 

Let us, to a first approximation, parametrize the various CDM 
density profiles, at very 
large radii (i.e. when we can completely neglect the baryonic component) as  
\begin{equation}
\rho_{\rm CDM}=\frac{\kappa~\bar{v}_c^2}{r^n}
\end{equation}
where $\kappa$ is a proper positive constant which depends on the particular 
profile that is chosen
~\cite{King:1962wi,Burkert:1995yz,Moore:1999gc,Navarro:1996gj,Navarro:2003ew,Salucci}. For example for many of the density 
profiles the slope is $n=3$ for large radii 
\cite{Burkert:1995yz,Moore:1999gc,Navarro:1996gj,Navarro:2003ew,Salucci}.    

In this case a positive energy density $\rho>0$ requires $n\ge 2$.
At this point let us focus on the case where $2 \le n<4$, since this gives 
rise to the typical slope of most of the 
density profiles studied in the literature.
Therefore we obtain  for $\rho(r)$ and $p_\parallel(r)$:
\begin{equation}
\label{rho-p_2<n<4}
\rho(r)=\bar{v}_c^2~\kappa\,\frac{n-2}{4-n}\frac{1}{r^n}\;,\quad\quad\quad
p_\parallel(r)=\bar{v}_c^2~\kappa\,\frac{2}{4-n}\frac{1}{r^n}\;.
\end{equation}
In particular,
\begin{itemize}
\item[1)]for $n=2$, we get
\begin{equation}
\label{rho-p_n=2}
\rho(r)=0,\quad\quad\quad
p_\parallel(r)=\rho_{\rm CDM}=\bar{v}_c^2~\kappa\,\frac{1}{r^2}\;,
\end{equation}
and for the relation between $\hat{r}_{(0)}$ and $\bar{r}$ one can choose, for example, $\hat{r}_{(0)}=\bar{r}=0$. 
In other words, for large radii we have that $\rho(r) \ll p_\parallel(r)$.
\item[2)]Also for $2<n<3$ one can choose $\hat{r}_{(0)}=\bar{r}=0$.
\item[3)]For $n=3$ 
\begin{equation}
\label{rho-p_n=2}
\rho(r)=\rho_{\rm CDM},\quad\quad\quad
p_\parallel(r)=\bar{v}_c^2~\kappa\,\frac{2}{r^3}\;,
\end{equation}
and, through Eq.~(\ref{initial-cond}), we have to impose that
\begin{equation}
\frac{M_{(1)}(\hat{r}_{(0)})-M_{{\rm CDM}\;(1)}(\bar{r})}{4 \pi}=
\ln{\left(\frac{\hat{r}_{(0)}}{\bar{r}}\right)}-2\;.
\end{equation}
Notice that the energy density profile is the same as the CDM one only for large radii so that  
$M_{(1)}(r)$ differs  from $M_{{\rm CDM}\;(1)}(r)$. 
\item[4)]In addition, for $3<n<4$, also through Eq.~(\ref{initial-cond}), 
we have to impose that
\begin{equation}
\frac{M_{(1)}(\hat{r}_{(0)})-M_{{\rm CDM}\;(1)}(\bar{r})}{4 \pi}=
\frac{\bar{r}^{3-n}}{n-3}-\frac{(n-2)}{(4-n)(3-n)}\hat{r}_{(0)}^{3-n}.
\end{equation}
\end{itemize}
Now let us focus where $2<n<4$. 
Starting from Eq.~(\ref{rho-p_2<n<4}) to express 
$p_\parallel=p_\parallel(\rho)$ we solve Eq.~(\ref{def-p_with_chi})
to recover the Lagrangian for the scalar field 
\begin{equation}
\label{rho-chi}
\rho(\chi)= - \mathcal{L}= k \chi^{\frac{n}{2(n-2)}}\;,\quad\quad\quad
p(\chi)=\frac{2k}{(n-2)}\chi^{\frac{n}{2(n-2)}}
\end{equation}
where $k$ is a positive integration constant.
We can see that, for this range of $n$, the exponent is larger 
than $1$; thus there are no problems with a possible instability
of the Lagrangian (see 
Refs.~\cite{ArmendarizPicon:2005nz,Babichev:2007dw,Rendall:2005fv}). 
Therefore, through the 
transformation $\rho \rightarrow \rho + \Lambda$ \;
$p_\parallel \rightarrow p_\parallel - \Lambda$, this
Lagrangian can be extended to describe a unified model of dark matter 
and dark energy.
Indeed, starting from the Lagrangian of the type (\ref{L-purelykessence}),
when $|X| \gg \hat{X}$ and if $k=g_n$, $\mathcal{L}$ takes precisely 
the form (\ref{rho-chi}).

Finally, we want to stress that this prescription does not apply only 
to the case of an adiabatic fluid, such as the one provided by 
scalar field with a purely kinetic Lagrangian, but it 
can be also used for more general Lagrangians 
$\mathcal{L}(\varphi,X)$.  


\section{Unified Dark Matter models with non-purely kinetic Lagrangians}

Let us consider more general Lagrangians of type 
$\mathcal{L}= \mathcal{L}(\varphi,X)$, with a non-canonical kinetic term, 
in order to find a UDM model with acceptable cosmological speed of 
sound.
In this case we have one more degree of freedom: 
the scalar field configuration 
itself. Therefore, we have to impose a new condition 
to the solutions of the equation of motion.
Ref.~\cite{Bertacca:2007ux}  
required that the Lagrangian of the scalar field is constant
along the classical trajectories. We want to know whether such a condition 
could be applied to the static spherically symmetric space-time
metric. We would also like to know what the behavior of the rotation 
velocity $v_c$ in the halo of a spiral galaxy is like for this class of models.
In the next subsections we will consider first a Lagrangian of the form
$\mathcal{L}=f(\varphi)g(X)$ and then a Lagrangian of the form  
$\mathcal{L}=g(X)-V(\varphi)$. In these cases, for simplicity, we will assume 
that $f(\varphi)$ and $V(\varphi)$ are positive. 

\subsection{Lagrangian of the type $\mathcal{L}=f(\varphi)g(X)$}

Let us write the Lagrangian in the form
$\mathcal{L}=f(\varphi)g(X)=-f(\varphi)g_s(\chi)$. 
Immediately we notice that the requirement of having a positive energy density
imposes that $g_s(\chi)$ is positive.
In this particular case the equation of motion (\ref{eq-motion}) becomes
\begin{equation} 
\label{eq-fg}
\fl
\frac{d}{d R}\left\{\ln{\left|2\chi\frac{d g_s(\chi)}{d\chi}-g_s(\chi)
\right|}\right\}+2\chi\frac{d g_s(\chi)}{d\chi}
\left[2\chi\frac{d g_s(\chi)}{d\chi}-g_s(\chi)\right]^{-1}
= -\frac{d\ln f(\varphi)}{d R}\;.
\end{equation}
Moreover from Eq.~(\ref{def-p_with_chi}) we obtain for 
$p_\parallel$
\begin{equation}
\label{def-p-fg}
p_\parallel=f(\varphi)g_s(\chi)
\left\{2\chi\frac{d \ln \left[g_s(\chi)\right]}{d\chi}-1\right\}\;.
\end{equation}
Following the procedure previously explained we impose the constraint
$\mathcal{L}=-\rho= -\Lambda$, i.e.
\begin{equation}
\label{lambda-fg}
f(\varphi)=\frac{\Lambda}{g_s(\chi)}\;, 
\end{equation}
which, inserted in the equation of motion (\ref{eq-fg}), allows to find the 
following general solution 
\begin{equation}
\label{gen_sol-fg}
\chi\frac{d \ln \left[g_s(\chi)\right]}{d\chi}=
\frac{k/2}{r^2\exp{(\alpha)}} \;, 
\end{equation}
where $k$ is a constant of integration.
Now, inserting Eqs.~(\ref{lambda-fg}) and (\ref{gen_sol-fg})
into the relation (\ref{def-p-fg}), we obtain
\begin{equation}
\label{gen_sol-L-const}
p_\parallel=\frac{\Lambda k}{r^2\exp{(\alpha)}}-\Lambda\;.
\end{equation}
Using this expression and considering the halo for $r>r_b$ and $M\gg M_*$,
we are finally able to get the expression for the rotation velocity 
\begin{equation}
\label{vc-fg}
v_c^2 \simeq \frac{\Lambda k/2}{\exp{(\alpha)}}-\frac{\Lambda r^2}{3}\;.
\end{equation}
If $\exp{(\alpha)} \approx$ const. this expression leads to a
flat rotation curve for all radii $r<r_{\rm max}$ such that 
$r_{\rm max}^2 \ll 3k/(2\exp{(\alpha)})$ and provided that 
the constant $k$ is positive. 
Therefore, in the future we will always neglect the second term 
in Eq.~(\ref{vc-fg}). 

It is important to stress that the results outlined in 
Eqs.~(\ref{lambda-fg})-(\ref{vc-fg}) give an efficient recipe to obtain 
a flat halo rotation curve within the UDM scenario. 
Once a Langrangian (i.e. $g(X)$) leading to a viable 
UDM model on cosmological scales is obtained by 
imposing the constraint $\mathcal{L}=-\Lambda$ 
(see Ref.~\cite{Bertacca:2007ux}), a flat rotation curve is 
guaranteed through Eq.(\ref{vc-fg}). There are however two important 
requirements that have to be satisfied. 
The function $g_s(\chi)$ must allow for a positive integration constant $k$ 
through Eq.(\ref{gen_sol-fg}), and the Lagrangian must 
satisfy the stability conditions discussed in
Refs.~\cite{ArmendarizPicon:2005nz,Babichev:2007dw,Rendall:2005fv}), 
which require $\partial \mathcal{L}/\partial X>0$ and 
$\partial \mathcal{L}/\partial X + 2X \partial^2 \mathcal{L}/\partial X^2 >0$ 
(so that the speed of sound is positive 
both in the cosmological setting and for the static solution).
     
In the second part of this subsection we will consider first 
a situation where $M_*=0$, in other words when our halo is composed only 
of the dark fluid, and then 
a situation where there is a non-negligible baryon contribution
in the inner part of the halo. 
 
\subsubsection{Case $M_*=0$: halo composed only of the dark fluid}

Starting from $\rho=\Lambda$ and 
$p_\parallel=\Lambda k/(r^2\exp{(\alpha)})-\Lambda$ we can 
explicitly calculate the value of $\exp{(\alpha)}$ and 
$\exp{(\beta)}$ through Eqs.~(\ref{eqE1})
and (\ref{eqE2}). Therefore, for $M_*=0$ we get
\begin{equation}
\label{exp-2beta-m*0}
\exp{(-2\beta)}=1-\frac{\Lambda r^2}{3} \;,
\end{equation}
\begin{equation}
\label{expalpha-m*0}
\exp{(\alpha)}=\frac{\Lambda k}{2}
\left\{\left(1-\frac{\Lambda r^2}{3}\right)^{1/2}
\left[\frac{2\kappa}{\Lambda k}-
\ln{\frac{\left(\frac{\Lambda}{3}\right)^{1/2}r}{1-\left(1-\frac{\Lambda r^2}
{3}\right)^{1/2}}}\right]+1\right\} \;,
\end{equation}
where $\kappa$ is a suitable positive integration constant. 
In particular,  
the value of $\kappa$ should be such that the term 
on the RHS of Eq.~(\ref{expalpha-m*0}) is positive, i.e.
\begin{equation}
\label{cond-positivity-vc}
\left(\frac{\Lambda}{3}\right)^{1/2}r >
\left[\cosh{\left(\frac{2\kappa}{\Lambda k}+1\right)}\right]^{-1}\;.
\end{equation}
It is very important to stress that, in this case, the weak energy conditions
are satisfied. In other words, through this prescription, we are able 
to evade the \emph{no-go} theorem derived in 
Ref.~\cite{DiezTejedor:2006qh}.

Using Eq.~(\ref{expalpha-m*0}) and Eq.~(\ref{vc-fg})
we can obtain the following expression for the circular velocity  
\begin{equation}
\label{vc-m*0}
v_c^2(r)=\left\{\left(1-\frac{\Lambda r^2}{3}\right)^{1/2}
\left[\frac{2\kappa}{\Lambda k}-
\ln{\frac{\left(\frac{\Lambda}{3}\right)^{1/2}r}{1-\left(1-\frac{\Lambda r^2}
{3}\right)^{1/2}}}\right]+1\right\}^{-1}.
\end{equation}
In order to have values of $v_c \sim 10^{-3}$ we must impose
that $2 \kappa/(\Lambda k) \sim 10^{6} \ll 3/(\Lambda r_{max}^2)$.
Imposing this condition we can obtain an approximately flat 
halo rotation curve.\\
  
A simple inspection of Eqs.~(\ref{expalpha-m*0}), 
(\ref{cond-positivity-vc}) and (\ref{vc-m*0}) shows an interesting 
property of our result.
There is a minimum radius $r_{min} \approx 
(\Lambda /3)^{-1/2}\left[\cosh{(2\kappa/(\Lambda k))}\right]
^{-1}$ required for the validity of (\ref{cond-positivity-vc}). 
Obviously it is necessary that
$r_{min} \ll r_{gal} ~ (\ll r_{max})$ where $r_{gal}$ is the typical radius
of our halo. 

\subsubsection{Case $M_*\neq 0$: non-negligible baryonic component
in the center of the halo\\}
In this subsection we assume that $r>r_b$. 
If $M\gg M_*$ we recover the same result of the previous
subsection; if $M_*\gg O(\Lambda r^3)$, using Eqs.~(\ref{eqE1})
and (\ref{eqE2}), we obtain 
\begin{equation}
\label{exp-2beta-m*}
\exp{(-2\beta)}\approx1-\frac{M_*}{4\pi r}
\end{equation}
\begin{equation}
\label{expalpha-m*}
\exp{(\alpha)}\approx\Lambda k
\left\{\left(1-\frac{M_*}{4\pi r}\right)^{1/2}
\left[\frac{\kappa_*}{\Lambda k}+
\cosh^{-1}\left(\frac{4\pi r}{M_*}\right)^{1/2}
\right]-1\right\}\;.
\end{equation}
where $\kappa_*$ is a suitable positive integration constant. In particular
it easy to see that $\kappa_*$ and $k$ (through $\exp{(\beta(r))}$) depend 
also on the value of $M_*$, since one is considering $r>r_b$.
Obviously, these functions exist only for $r>r_*>M_*/(4\pi)$, having defined
$r_*$ as the value of the radius for which $\exp{(\alpha(r_*))}=0$.
In this case, using the approximate relation (\ref{v_c-2}), $v_c$ reads
\begin{equation}
\label{vc_m*}
v_c^2\approx\frac{1}{2}\left\{\left(1-\frac{M_*}{4\pi r}\right)^{1/2}
\left[\frac{\kappa_*}{\Lambda k}+
\cosh^{-1}\left(\frac{4\pi r}{M_*}\right)^{1/2}\right]-1\right\}
^{-1}+\frac{M_*}{8\pi r}\;.
\end{equation}
To have halo rotation velocities $v_c \sim 10^{-3}$ for $r\gg r_*$, we 
need to impose $2 \kappa_*/(\Lambda k) 
\sim 10^{6}$. One can see that this condition leads to   
$r_*\approx M_*/(4\pi)$. Moreover, also in this case we have
a minimum radius $r_{min}$ such that $v_c^2(r_{min})=1$.
Starting from Eq.~(\ref{v_c}) we get
\begin{equation}
r_{min}\approx \frac{M_*}{2\pi}> r_*\;.
\end{equation}

\subsection{Lagrangian of the type $\mathcal{L}=g(X)-V(\varphi)$}

In this subsection we briefly discuss Lagrangians of the type 
$\mathcal{L}=g(X)-V(\varphi)$. 
Let us rewrite $\mathcal{L}$ as 
$\mathcal{L}=-[g_s(\chi)+V(\varphi)]$.
In order to have $\rho>0$ we impose that $g_s(\chi)>0$. For these Lagrangians
the equation of motion (\ref{eq-motion}) becomes
\begin{equation}
\label{eq-motion-g-v}
\chi\frac{d g_s(\chi)}{d\chi}\left\{
\frac{d\ln\left[\chi\left(\frac{d g_s(\chi)}{d\chi}\right)^2
\right]}{d R}+2\right\}=\frac{d\ln V(\varphi)}{d R} \;.
\end{equation}
Requiring that the Lagrangian of the scalar field is constant
along the classical trajectory, i.e
\begin{equation}
V(\varphi)=-g_s(\chi)+\Lambda \;, 
\end{equation}
from the Eq.~(\ref{eq-motion-g-v}) we get 
\begin{equation}
\label{gen_sol-g-v}
\chi\frac{d g_s(\chi)}{d\chi}=\frac{k/2}{r^2\exp{\alpha}}
\end{equation}
where $k$ is a positive constant.\\
Now, inserting Eq.~(\ref{gen_sol-g-v}) into Eq.~(\ref{def-p_with_chi})
we obtain the same expressions for $p_{\parallel}$ and $v_c$
that we obtained in the last subsection i.e. 
Eqs.~(\ref{gen_sol-L-const}) and (\ref{vc-fg}), respectively. 

\section{Conclusions}

In this paper we have investigated static spherically symmetric
solutions (``dark halos'')
of Einstein's equations for a scalar field with non-canonical 
kinetic term. 
Assuming that the scalar field depends only on the radius, we 
studied Unified Dark Matter models with purely kinetic Lagrangians.
In particular, we obtained a purely kinetic
Lagrangian which allows simultaneously to produce flat halo rotation curves 
and to realize a unified model of dark matter and dark energy 
on cosmological scales. Moreover, we gave a prescription to obtain UDM model 
solutions that have the same rotation curve $v_c(r)$ as a CDM model with a 
specified density profile. 
Next, we considered a more general class of Lagrangians 
with non-canonical kinetic term. 
In this case we have one more degree of freedom (the scalar field 
configuration itself) and we need to impose one more constraint. To this aim,
we required that the Lagrangian is constant, $\mathcal{L}=-\Lambda$ 
along the solutions of the equation of motion.
We have studied whether this condition 
can be applied to the static spherically symmetric space-time 
metric and what the behavior of $v_c(r)$ is for this class of models.

Let us finally stress that these solutions allow for the possibility 
to find suitable Lagrangians that describe with a single fluid 
viable cosmological and static solutions. 

\section*{Acknowledgments}
We thank Massimo Pietroni, Antonaldo Diaferio, Paolo Salucci and 
Shinji Tsujikawa 
for useful discussions.

\vskip 1cm
\appendix
\setcounter{equation}{0}
\def\theequation{A.\arabic{equation}}
\vskip 0.2cm
\section{Spherical collapse for purely kinetic scalar-field Lagrangians}

Let us assume a flat, homogeneous Friedmann-Robertson-Walker 
background metric i.e.
\begin{equation}
ds^2=-dt^2+a(t)^2\delta_{ij} dx^i dx^j\, ,
\end{equation}
where $a(t)$ is the scale factor, $\delta_{ij}$ denotes the unit tensor. 
Moreover, for this particular case, the Hubble parameter $H$ is a function 
only of the UDM fluid $H^2 = \rho/3$.\\
Now let us consider a top-hat spherical overdensity with the purely kinetic
model with the Lagrangian $\mathcal{L}= -\Lambda + g_n (X-\hat{X})^n $ and
with $g_n>0$ \cite{Bertacca:2007ux}. For this particular case
within the overdense region we have a single dark fluid ondergoing 
spherical collapse, which is described by the following equation 
\begin{equation}\label{qr2_lammda=0}
\frac{\ddot R}{R} = -\frac{1}{6} \left(\rho_{R} + 
3 p_{R} \right)
\end{equation}
where $R(t)$, $\rho_{R}$ and $p_{R}$ are respectively the scale-factor,
pressure and energy density of the overdense region and where the dot denotes 
differentiation w.r.t. the cosmic time $t$.\\  
Now, $\rho_{R}$ and $p_{R}$ are defined by the following expressions 
\cite{Bertacca:2007ux}
\begin{equation}
\label{rho_overdensity}
\rho_{R} = \Lambda + 2 n g_{n} \hat{X} (X_R - \hat{X})^{n-1} + 
(2n-1) g_{n} (X_R-\hat{X})^{n}
\end{equation}
\begin{equation}
\label{p_overdensity}
p_R= g_R =- \Lambda+g_{n}(X_R-\hat{X})^{n}
\end{equation}
with $X_R=X(R)$ a function of time. 

The equation of motion is
\begin{equation}
\label{collapse_lambda=0}
 \left(\frac{\partial g_R}{\partial X_R}+
2X\frac{\partial^2 g_R}{\partial X_R^2}\right)\frac{dX_R} {dN_R}
+ 3 \left(2 X_R \frac{\partial g_R}{\partial X_R}\right) = 0 \;. 
\end{equation}
where $dN_R=dR/R$. The solution of Eq.~(\ref{collapse_lambda=0})
(for $\partial g_R/\partial X_R , X_R \neq 0$ ) is
\begin{equation}
\label{sol_coll}
X_R\left(\frac{\partial g_R}{\partial X_R}\right)^{2} =
  k_R R^{-6} \end{equation}
where we can choose $k_R=R_{ta}^{6} \left[ X_R 
\left(\frac{\partial g_R}{\partial X_R} \right)^{2} \right]_{ta}$, with 
$R_{ta}$ the value of $R$ at turnaround.
Replacing  Eq.~(\ref{p_overdensity}) in Eq.~(\ref{sol_coll}) we find
\begin{equation}
\label{sol_coll2}
X_R \; \left[n g_n (X_R-\hat{X})^{n-1}\right]^2=  k_R R^{-6}
\end{equation}
Using now the explicit expressions for $\rho_R$ and $p_R$ we arrive at the 
following set of equations 
\begin{eqnarray}
\label{collapse_gen-eq}
\frac{\ddot R}{R}  = -\frac{1}{3} \left[- \Lambda + 
n g_{n} \hat{X} (X_R-\hat{X})^{n-1} + (n+1) g_{n} 
(X_R-\hat{X})^{n}\right] \\
(X_R-\hat{X})^{2n-1} + \hat{X} (X_R-\hat{X})^{2(n-1)} = 
\frac{k_R} {n^2 g_{n}^{2}} R^{-6}.
\end{eqnarray}

For $(X_R-\hat{X})/\hat{X} \ll 1$ Eq.~(\ref{collapse_gen-eq}) 
becomes
\begin{equation}
\label{collapse_eq_Xc_near_Xco}
\frac{\ddot R}{R} = - \frac{1}{3} \left\{- \Lambda +  n g_{n} 
|X_{R_{ta}}-\hat{X}|^{n-1} (X_{R_{ta}}\hat{X})^{\frac{1}{2}} 
\left(\frac{R} {R_{ta}}
\right)^{-3} \right\} 
\end{equation}

We can now write all the equations that describe 
the spherical collapse
\begin{eqnarray}
\label{eq_final_a}
\left(\frac{\dot a}{a} \right)^{2}  & = 
& \frac{1}{3} \left(\rho_\Lambda +  \rho_{\rm DM}\right) \\
   \rho_\Lambda & = & \Lambda \\
\label{eq_final_rho_k_DM}
       \rho_{\rm DM} & = 
& 2 n g_n |X_{ta}-\hat{X}|^{n-1} (X_{ta}\hat{X})^{\frac{1}{2}} 
\left(\frac{a} {a_{ta}}\right)^{-3} \\
\label{eq_final_R}
\frac{\ddot R}{R} & = & - \frac{1}{6} \left(\rho_{R_{DM}} 
- 2\rho_{R_{\Lambda}}\right) \\
\rho_{R_{\rm DM}} & = & 2 n g_{n} |X_{ta}-\hat{X}|^{n-1} 
(X_{R_{ta}}\hat{X})^{\frac{1}{2}} \left(\frac{R} {R_{ta}}\right)^{-3}
\end{eqnarray}
where $a_{ta}=a(t_{ta})$. 

Following now the same procedure of Ref.~\cite{Wang-Steinhardt} 
we can define $x$ and $y$ 
\begin{eqnarray}
\label{}
 x & \equiv & \frac{a} {a_{ta}} \\
 y & \equiv & \frac{R} {R_{ta}} \; .
\end{eqnarray}
In this way we can redefine $\rho_{\rm DM}$ and $\rho_{R_{\rm DM}}$ such that
\begin{eqnarray}
\label{eq_final_xy}
\rho_{\rm DM} & = & \frac{3 H_{ta}^{2}\Omega_{\rm DM}(x=1)} 
{x^{3}} \\
 \rho_{R_{\rm DM}} & = & \zeta \frac{3 H_{ta}^{2}\Omega_{\rm DM}(x=1)} 
{y^{3}} 
\end{eqnarray}
where $\Omega_{\rm DM}$ is the (k-essence) dark matter density parameter, 
and $\zeta = (\rho/\rho_{\rm DM})|_{x=1}$. Then Eqs.~(\ref{eq_final_a}) 
and (\ref{eq_final_R}) become
\begin{eqnarray}
\label{}
        \frac{d x}{d \tau}  & = 
& \left(x \Omega_{DM}(x) \right)^{- \frac{1}{2}} \;, \\ 
\frac{d^{2} y}{d \tau^{2}}  & = & -\frac{1}{2y^{2}}\left[\zeta - 2y^{3} 
K_{\Lambda} \right] \;, \\
             \Omega_{DM}(x)  & = & \left(1- 
\frac{1-\Omega_{DM}(x=1)}{\Omega_{DM}(x=1)} 
x^{3}\right)^{-1} \;, 
\end{eqnarray}
where $ d \tau = H_{ta} \sqrt{\Omega_{DM}(x=1)}$ and 
$ K_{\Lambda} = \rho_\Lambda/[3H_{ta}^2 
\Omega_{DM}(x=1)]$. 

Defining $U$ as the potential energy of the overdensity 
and using energy conservation between virialization and turnaround, 
\begin{equation} 
\left[ U+\frac{R}{2}\frac{\partial U}{\partial R}  
\right]_{vir}  =  U_{ta} \label{ec} \;, 
\end{equation}
we obtain 
\begin{equation}
\label{}
(1+q)y-2qy^{3} = \frac{1}{2}
\end{equation}
where
\begin{equation}
\label{}
q = \left(\frac{\rho_{\Lambda}}{\rho}\right)_{y=1} = 
\frac{K_{\Lambda}}{\zeta} \;, 
\end{equation}
in full agreement with Ref.~\cite{Lahav}.

\end{document}